\documentclass[conference,onecolumn,10pt,letterpaper]{IEEEtran}
\IEEEoverridecommandlockouts
\usepackage{cite}
\usepackage{amsmath,amssymb,amsfonts}
\usepackage{algorithmic}
\usepackage{graphicx}
\usepackage{textcomp}
\usepackage{xcolor}
\def\BibTeX{{\rm B\kern-.05em{\sc i\kern-.025em b}\kern-.08em
    T\kern-.1667em\lower.7ex\hbox{E}\kern-.125emX}}

\usepackage{listings}

\definecolor{vgreen}{RGB}{104,180,104}
\definecolor{vblue}{RGB}{49,49,255}
\definecolor{vorange}{RGB}{255,143,102}

\lstdefinestyle{verilog-style}
{
	language=Verilog,
	basicstyle=\small\ttfamily,
	keywordstyle=\color{vblue},
	identifierstyle=\color{black},
	commentstyle=\color{vgreen},
	numbers=left,
	numberstyle=\tiny\color{black},
	numbersep=10pt,
	tabsize=8,
	moredelim=*[s][\colorIndex]{[}{]},
	literate=*{:}{:}1
}
\usepackage{pgf-pie}  
\usepackage{pgfplots}
\usepackage{caption}
\usepackage{subcaption}
\usepackage{geometry}
\usepackage{comment}
\begin{document}

\title{Automated Formal Verification of a Highly-Configurable Register Generator}

\author{\IEEEauthorblockN{Shuhang Zhang}
\textit{Infineon Technologies AG}\\
Neubiberg, Germany \\
Shuhang.Zhang@infineon.com
\and
\IEEEauthorblockN{Bryan Olmos}
\textit{Infineon Technologies AG}\\
Neubiberg, Germany \\
Bryan.Olmos@infineon.com
\and
\IEEEauthorblockN{Basavaraj Naik}
\textit{Infineon Technologies AG}\\
Neubiberg, Germany \\
Basavaraj.Naik@infineon.com
}

\maketitle

\begin{abstract}
	\textbf{Registers in IP blocks of an SoC perform a variety of functions, most of which are essential to the SoC operation. The complexity of register implementation is relatively low when compared with other design blocks. However, the extensive number of registers, combined with the various potential functions they can perform, necessitates considerable effort during implementation, especially when using a manual approach. Therefore, an in-house register generator was proposed by the design team to reduce the manual effort in the register implementation. This in-house register generator supports not only the generation of register blocks but also bus-related blocks. Meanwhile, to support various requirements, 41 generation options are used for this generator, which is highly-configurable. From the verification perspective, it is infeasible to achieve complete verification results with a manual approach for all options combinations. Besides the complexity caused by configurability, the register verification is still time-consuming due to two widely recognized issues: the unreliability of specifications and the complexity arising from diverse access policies. To deal with the highly-configurable feature and both register verification issues, we propose an automated register verification framework using formal methods following the Model Driven Architecture (MDA). Based on our results, the human effort in the register verification can be reduced significantly, from 20Person-Day (20PD) to 3PD for each configuration, and 100\% code coverage can be achieved. During the project execution, eleven new design bugs were found with the proposed verification framework.}
\end{abstract}
\section{Introduction}
With more functions integrated into SoCs, the number of registers used to configure and control the behavior also increases significantly. These registers are critical components and their proper functioning ensures that manufactured chips operate as intended. Although the implementation difficulty of registers is low, implementing numerous registers with different specifications manually is usually inefficient and error-prone. More specifically,
a register generator can offer several benefits over writing register code manually. One of the most significant advantages is that it can significantly reduce the development time. This is because a register generator can effortlessly create code for hundreds or thousands of registers within seconds.
Additionally, a register generator can ensure consistency in the naming and definition of registers across the entire design. 
Moreover, a register generator is also more flexible than manual coding, as they can be configured to generate code optimized for different design requirements. Finally, they can help reduce human error, which can be a significant problem with manual coding~\cite{pearce2020dave,Thakurverilog}.
Therefore, a register generator is always desired in the industry, but the generator based on the large language model~\cite{gpt4} has not been utilized for the HDL code generation \textcolor{black}{for automotive applications} because of functional safety considerations~\cite{palin2011iso}.

Motivated by the promising advantages and considering the functional safety perspective, an in-house generator is created, which follows the Model Driven Architecture~(MDA)~\cite{mda} and is based on code template definitions~\cite{ecker2016introducing,devarajegowda2021model}. This generator supports not only the generation of various types of registers following diverse specifications but also bus interface generation. In addition, 41 generation options are utilized by the generator to support various requirements. The reason why we develop an in-house register generator instead of adopting a general-purpose register generator is some highly-customized specifications required by the internal departments.

With this in-house tool utilized widely, the design effort can be reduced significantly, but the verification of this register generator tends to be a great challenge for the verification engineer. As thousands of registers can be generated in seconds, these registers need to be verified completely to ensure that they are working correctly, which makes the manual approach infeasible for it. Furthermore, the increasing complexity of SoCs means that more special functions should be considered during the register verification. Meanwhile, the progress in automated register verification falls behind when compared with the significant progress in the development of register generator. The traditional approach used in our department is based on  the Universal Verification Methodology~(UVM).
However, based on our experience, the effort spent in adapting the testbench and UVM setup for each option configuration is huge and it is impossible to handle a large number of configurations in a reasonable time. Besides the configurable feature of the register generator, there are always some  gaps in the simulation-based verification, because of the time-consuming and error-prone UVM testbench adaptation caused by the arising complexity  from diverse access policies~\cite{chen2017challenges}. Therefore, we would like to find an alternative approach to improve the verification efficiency and eliminate this gap for the register verification.

In this work, to address the verification challenges brought by the highly-configurable register generator, we propose a novel verification framework. More specific, the contributions of this work include:
\begin{itemize}
\item	We analyze our in-house register generator and propose to use the formal approach to verify the generated RTL code.
\item	We propose a novel verification flow following the MDA~\cite{mda} to generate the property files based on harmonized specifications automatically. These property files can be used by the formal tools naturally.
\item	Our results show that the manual effort for the register verification can be reduced significantly, from 20Person-Day (20PD) to 3PD for one configuration. In addition, our formal framework detected eleven bugs for the register generator which has been verified using the simulation-based approach.
\end{itemize}

The rest of this paper is organized as follows. In Section~\ref{sec:background}, we introduce our in-house register generator and highlight the motivation for the automated formal verification framework. The proposed formal verification framework will be elaborated in Section~\ref{sec:framework}, and the verification results will be presented in Section~\ref{sec:results}. At last, we conclude this work in Section~\ref{sec:conclusion}.

\section{Background}\label{sec:background}
In this section, we would like to explain some background concepts used by the proposed framework and highlight the challenges for the register verification.
Firstly, we introduce how register specification are defined in this work. 
Then more details regarding the in-house register generator are presented. Its configurable features are also introduced, which makes it flexible for various specifications. Afterwards, we give some basic knowledge about the MDA. Lastly, we explain the motivation for the automated formal verification framework.

\subsection{XML Specification File}
The unreliability of specifications is always one main issue for the register implementation and verification, as it is prevalent for some special accesses are omitted from the specifications. This omission necessitates significant effort of the verification engineer, who has to undertake a lot of trials to understand the correct behavior of these registers~\cite{chen2017challenges}. A harmonized specification format is always desirable, so many candidates have been under discussion internally for a quite long time. Finally, the XML format was chosen as the default specification format for the register specification. Within the XML file, all required attributes  and custom features can be defined.
In this way, we have achieved a harmonized specification for both design and verification teams. 

\subsection{Register Generator Overview}\label{sec:background:bsg}

\begin{figure}
	\centering
	\includegraphics{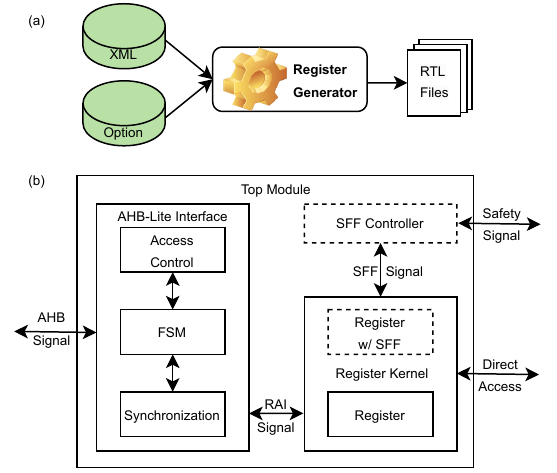}
	\caption{(a) Overview of the in-house register generator. (b) Simplified diagram of generated RTL blocks.}
	\label{fig:bsg_gen}
\end{figure}

In Fig.~\ref{fig:bsg_gen}(a), the working flow of our in-house register generator is illustrated. This generator has two inputs, XML file and generation options. The first input is an XML file where basic features of specifications are defined, e.g. Address Value and Access Policy.
The other input are the options used by the generator, which are usually defined in a Makefile. These options can also be considered additional specifications, which cannot be defined in the XML file easily. As we have 41 different options, it is not a good idea to explain all of them in this paper, so we present three example options shown as follows:
\begin{itemize}
	\item regUnrollAHB decides if AHB signals are unrolled or rolled, e.g. haddr\_i or sx\_ahb\_lite\_slave.haddr,
	\item regAsync decides if an asynchronous clock is used by the register kernel,
	\item regBusClock can specify some registers still work at the bus clock only when regAsync is set True.
\end{itemize}

After presenting the basic working flow of the register generator, we would like to give more details about the generated RTL blocks. As mentioned before, this generator can generate not only the register kernel but also the bus interface block as shown in Fig.~\ref{fig:bsg_gen}(b). The bus interface in this block acts as a slave role following the AHB protocol and takes AHB signals from the Master. Afterwards, the bus interface will transform the received AHB signals into Register Access Interface~(RAI) signals, which follows an internal protocol. Inside the bus interface, the Finite State Machine~(FSM), Access Control and Synchronization blocks are generated. Regarding the configurable feature, we would like to use the synchronization block as an example. When an asynchronous clock should be used for the register kernel~(configured via regAsync),  asynchronous FIFOs inside the synchronization block shall be generated to handle the data transferring between the bus and register clock domains. Apart from the bus interface block, all registers defined in the XML file are implemented inside this register kernel. In addition, our in-house generator also supports the integration of safety IPs into the register implementation. For example, the normal Flip Flop~(FFs) can be replaced with Safety Flip Flops~(SFFs) to comply with automotive safety standards, e.g. ISO 26262~\cite{ISO26262-6,palin2011iso}. If any SFF register is implemented in the register kernel, a SFF controller shall be instantiated in the top module, which controls the test for SFF instances and collects the alarm signals from the register kernel~\cite{busch2016automated}.

\subsection{MoT, MoP and MoV}
According to the MDA instructions, three models at different abstraction levels shall be used: Computation Independent Model (CIM), Platform Independent Model (PIM), and Platform Specific Model (PSM)~\cite{mda}. Each model-to-model transformation adds more details to create a less-abstract model. These model names are customized for the property generation flow. CIM, PIM and PSM are replaced with Model of Things~(MoT), Model of Property~(MoP) and Model of View~(MoV) respectively~\cite{devarajegowda2021model}. With this flow, the MoT layer is used to associate the XML specification into a metamodel, which can be used as an entry point for the design or verification. Then properties are defined in MoP layer. Afterwards, the property format is defined in MoV layer. At last, the property files can be generated as intended.

\subsection{Verification Challenges}
As mentioned in the previous sections, our current verification environment encounters two primary challenges, which are listed as follows:
\begin{itemize}
\item 	Configurability: Our in-house generator is highly configurable to support various features. However, the simulation-based approach requires usually 20PD for changing the testbench, debugging and making regression clean with each configuration. In this scenario, it is infeasible to verify a large number of configurations exhaustively.
\item 	Diverse access policies: As additional functions are integrated into an SoC, more registers are used. Typically, these registers are designed to support increasingly complex access policies. For example, registers can support external and internal accesses with different kinds of action behaviors and protection methods~\cite{chen2017challenges}.
\end{itemize}
In this work, to deal with above issues in the generator verification and achieve exhaustive results, we have proposed an automated formal verification framework with harmonized specifications. 

\section{framework}\label{sec:framework}
As mentioned in previous sections, the highly-configurable feature of the register generator along with two well-known register verification issues: unreliable specifications and diverse access policies makes the verification of this register generator a challenging task. However, with the harmonized specification format~(XML File), the unreliable specification issue was resolved. Therefore, we mainly address the remaining two issues with our proposed verification framework based on the MDA. In this framework, the verification process is broken down into several steps considering these abstraction models and detailed introduction is presented for each step in this section. 

\begin{figure}
	\centering
	\includegraphics{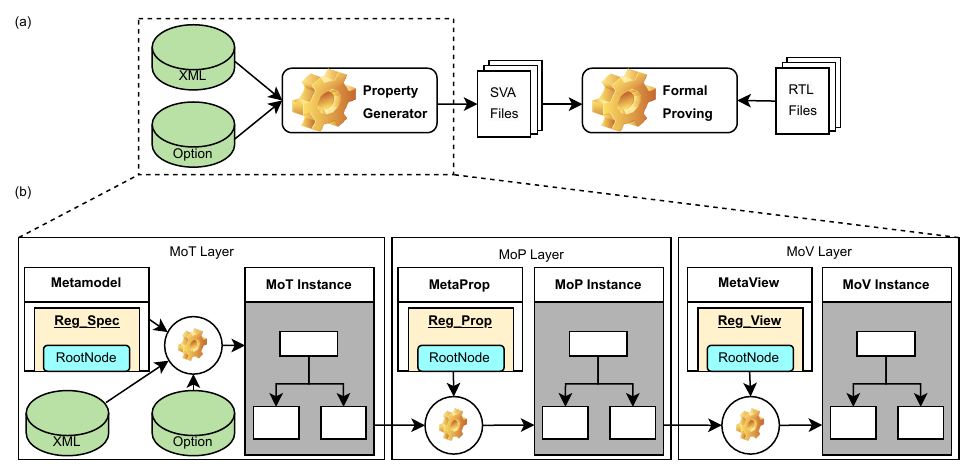}
	\caption{Proposed property generation flow for the register generator.}\label{fig:prop_flow}
	\vspace{-10pt}
\end{figure}

\subsection{Register Generator Analysis}
Before discussing the automated property generation, we would like to firstly point out that the formal approach is a good fit for the generator verification. As we know, the formal approach can check the state space exhaustively and the main issue of formal approach is the state explosion, which makes the checking inconclusive~\cite{clarke2012model}. However, regarding the verification of the register generator, the complexity of the generated register designs is usually low, because the intended behavior takes a small number of clock cycles. For instance, if the register kernel also operates at the bus clock, the read/write behaviors via the bus interface can be finished within two bus cycles. Even though the asynchronous register clock is utilized, the read/write operations can also be implemented in a reasonable time~(Some bus and register clock cycles depending on the FIFO depth utilized). As mentioned before, our in-house generator has 41 different generation options impacting the final generated RTL codes. We have analyzed the functionalities of all options with the designer together and all intended behaviors of all potential designs should be finished in a small number of clock cycles. Therefore, the formal approach was decided for the generator verification and an automated property generation flow will be the main focus of following sections.

In addition, verifying all combinations of options is infeasible in a reasonable time. The verification of all combinations is also not our target, because 
some are in conflict with each other. Therefore, to accelerate the verification progress, these generation options are classified with the design team. 
These options are classified 
into two parts, independent and dependent options. For the independent options, they shall be verified individually, e.g. the regUnrollAHB option rolling/unrolling AHB interface signals. Dependent options will be verified with valid combinations based on their dependency. For example, the regBusClock option allows some registers to operate still at the bus clock when an asynchronous clock is specified for the register kernal~(regAsync is set True).

\subsection{Specification Extraction in MoT Layer}
\begin{figure}
	\centering
	\includegraphics{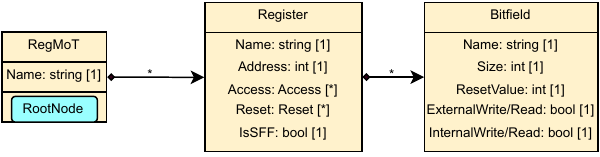}
	\caption{Simplified MoT model for specification extraction.}\label{fig:mot}
	\vspace{-10pt}
\end{figure}

Regarding the property generation, an internal specification representation is required for our in-house Python-based generation flow. Therefore, the information of XML specifications and configurable option values will be extracted automatically, Then MoT instances will be constructed accordingly at the MoT layer shown in Fig.~\ref{fig:prop_flow}(b), which are used as the input for the following MoP layer. As the MoT layer requires verification engineer to construct the MetaModel for the specification extraction, we present a simplified MetaMode in Fig.~\ref{fig:mot}. In this model, we need to define a RootNode firstly. Under this RootNode, multiple registers can be implemented and under each register multiple bitfields are allowed. In the register class, we can define multiple attributes, e.g., the address and access policy. Bitfield-related attributes are defined in the bitfield class, e.g., the size and read/write permission. Afterwards, this MoT model can be used to extract the specifications from the XML file and instantiated concrete MoT instances.

\subsection{Property Definition in MoP Layer}
With all internal specifications available, either from the XML file or generation options, the property class can be drafted in the MoP layer, also shown in Fig.~\ref{fig:prop_flow}(b). Afterwards, various kinds of properties are generated based on the classes defined using Python in this step, which makes this step the most important in the whole flow.

Some example property classes are listed as follows and the example properties will be shown in the  section~\ref{sec:framework:mov}.
\begin{itemize}
\item	Bus Protocol: As the AHB protocol can be verified easily using the Formal VIP~(FVIP), we focus on the decoding process from AHB to RAI signals with this property class when no access violations are triggered. 

\item	External Read/Write: We define the access via the bus interface as external access. With this property class, we check the normal read and write operations via the bus interface. No access violations are considered in this case, because all violations shall be verified using an individual property class.

\item	Access Violations: As diverse access policies are supported by the generated registers, most of them are protected with various signals and different error types can be defined accordingly in the XML file. With this property class, we will examine all violation scenarios in the protected registers to check correct error behavior.

\item	Internal Read/Write Operations: We define the register access form other RTL blocks as internal access, because the generated registers can be accessed from other RTL blocks directly instead of via the bus interface based on the XML file, so we check this internal read and write operations for the generated RTL code using this property class.

\item	Dummy Write: With this property class, we check if the register values can be modified when no write signals are enabled based on the external and internal specifications.

\item	Integrated Safety IP: The most important safety IP used by the generator is SFF, which is a formally pre-verified safety IP. Therefore, its functionality is not verified at the same level done by the SFF verification team and only basic behaviors are checked. In the register verification, we focus more on the integration checks, e.g. the connectivity and parameter of SFF components.
\end{itemize}


\begin{figure}
	
	\begin{subfigure}[b]{.9\linewidth}
		\vspace{-10mm}
		\includegraphics[scale=0.75]{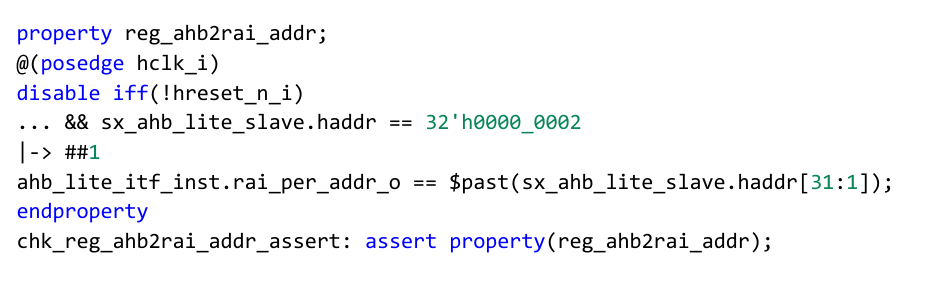}
		\vspace{-7mm}
		\subcaption{One simplified property to check the address decoding process from AHB to RAI signals without access violations (without bus errors).}
	\end{subfigure}
	
	\begin{subfigure}[b]{.9\linewidth}
		\includegraphics[scale=0.75]{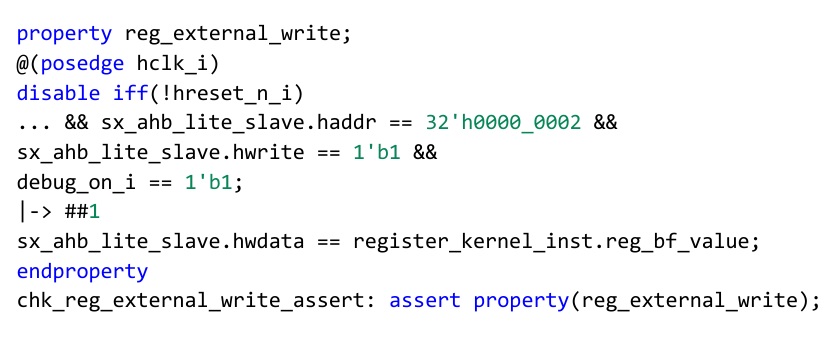}
		\vspace{-6mm}
		\subcaption{One simplified property to check the external write behavior without access violations (debug\_on\_i signal is high).}
	\end{subfigure}
	
	\begin{subfigure}[b]{.9\linewidth}
		\includegraphics[scale=0.75]{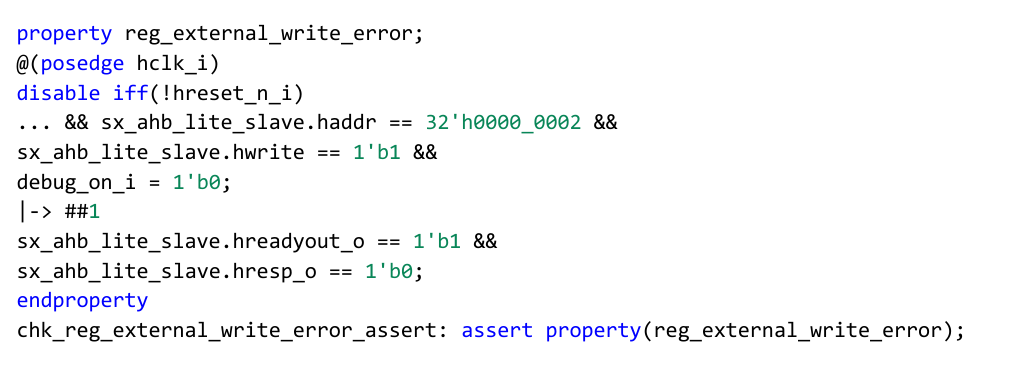}
		\vspace{-6mm}
		\subcaption{One simplified property to check the access violation behavior (debug\_on\_i is low and hwrite is high), so a bus error is expected.}
	\end{subfigure}
	
	\begin{subfigure}[b]{.9\linewidth}
		\includegraphics[scale=0.75]{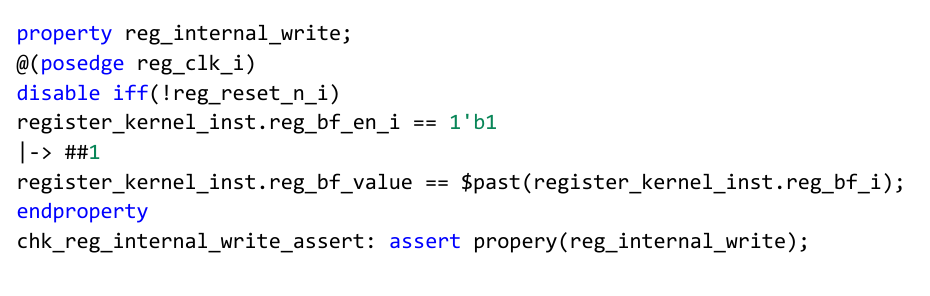}
		\vspace{-6mm}
		\subcaption{One simplified property to check the internal write behavior, where reg\_bf\_en\_i and reg\_bf\_i are primary inputs for the generated design.}
	\end{subfigure}
	
	\begin{subfigure}[b]{.9\linewidth}
		\includegraphics[scale=0.75]{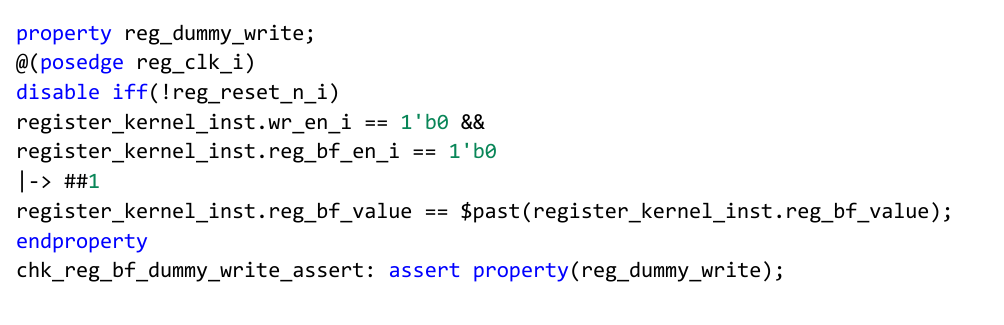}
		\vspace{-6mm}
		\subcaption{One simplified property to check the dummy write behavior when the external write enable~(wr\_en\_i) and internal write enable~(reg\_bf\_en\_i) are low.}
	\end{subfigure}
	
	\begin{subfigure}[b]{.9\linewidth}
		\includegraphics[scale=0.75]{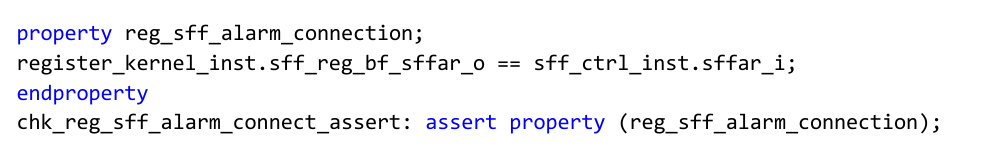}
		\vspace{-4mm}
		\subcaption{One simplified property to check the alarm connectivity between the safety register in the register kernel and SFF controller.}
	\end{subfigure}
	
	\caption{Example properties.}\label{fig:example_prop}
\end{figure}

\subsection{MoV for Property Generation}\label{sec:framework:mov}
All required property classes defined in the previous step will be mapped to target language models using the corresponding MoV to generate the intended property files, also shown in Fig.~\ref{fig:prop_flow}(b). All property classes defined in the last section will be applied for each register (and its bitfields if applicable) based on its specifications.
In addition, the SVA format is used in this work because it is the most popular format in the formal verification which is supported by almost all formal tools. 
Meanwhile, we would like to point out that our framework supports not only the SVA format but also other property formats, which makes it flexible for different formal tools.

As all properties defined in the last step shall be generated in this step, we list some example properties for an example register in this section.  In the following simplified properties shown in Fig.~\ref{fig:example_prop}, we assume the register has only one bitfield and we use \textit{reg\_bf} to represent the register and bitfield name declared in the register kernel. We also assume both external read and write are allowed only when the debug signal is high. Otherwise this register is read-only and a bus error occurs if writing via the bus interface. Regarding the internal access, we assume this register has both read and write operations. In the example design, we also assume AHB signals are rolled~(regUnrollAHB=False) and an asynchronous clock is not used~(regAsync=False). In addition, we omit some preconditions due to the page limit in Fig.~\ref{fig:example_prop}(a), (b) and (c).

In Fig.~\ref{fig:example_prop}(a), we give a simple example to check if the AHB address can be transformed into RAI signal correctly. Based on the RAI protocol, the AHB address shall be sliced after the bus block.
In Fig.~\ref{fig:example_prop}(b), one example property is used to check the  external write behavior when the write protection is fulfilled.
In Fig.~\ref{fig:example_prop}(c), we check if bus error can be generated when the access violation is triggered. By the way, we have individual properties to check the error signal should take two cycles.
In Fig.~\ref{fig:example_prop}(d), we check the internal write for this register.
In Fig.~\ref{fig:example_prop}(e), when no write enable signals, the register data should be the same.
In Fig.~\ref{fig:example_prop}(f), if this register is implemented with the SFF, some specifically-designed connectivity and parameter properties are applied to it, because these connectivity or parameter issues will not affect the normal behaviors of registers, which cannot be detected using the normal read/write properties.


\subsection{Formal Proving}
After generating all properties, these properties along with generated RTL files shall be loaded into a commercial formal tool to figure out if there are design issues. Afterwards, a coverage collection tool shall be used to check the completeness of generated properties.

\section{results}\label{sec:results}
After elaborating all steps of the proposed verification framework, we would like to present some results to demonstrate the advantages of this work over our old approach~(UVM-based verification) from different perspectives. In this work, we use three industry standard tools in this work, including a formal property checking tool, a coverage collection tool, and a clock  domain crossing checking tool.

\subsection{Effort Reduction}
 As mentioned in the previous section, the generator verification needs a one-time effort to draft all property classes. Afterwards, all properties with different configurations can be generated automatically. The one-time effort spent for this generator is around 80PD to achieve complete verification results for different generation options with a superset XML file. Afterwards, the effort for one configuration is reduced to 3PD, including the formal property checking, debugging, coverage discussion with designers and documentation preparation. When compared with the traditional approach~(20PD), it saves around 85\% effort for each configuration. We would like to highlight that the one-time effort is certainly worthwhile when considering the huge effort saved for verifying a large number of configurations. 
 
 \subsection{Coverage Improvement}
 Besides the effort reduction, we also compare the coverage results between our old UVM-based approach and this automated formal framework using the real specifications~(XML+Option  Configurations) from one automotive project. 
 In Fig.~\ref{fig:results}(a), the functional and code coverage results are presented. The functional coverage results of both approaches reach at $100\%$, but the code coverage with waivers still differs a lot. The detailed code coverage can be found in Fig.~\ref{fig:results}(b). The same waiver files, which are confirmed by the design team, are applied for both approaches. 
 In addition, we also analyze the low code coverage in our UVM-based approach and we found the verification gap is caused by the diverse access policies and customized register features.
 As mentioned before, our UVM-based approach requires a large time in adapting the setup when a new configuration is requested to be verified, so the UVM-based approach can only handle some basic features and cannot verify all specified features within a short time. What we want to emphasize here is that this comparison does not mean the UVM-based approach cannot achieve complete verification results, but our UVM-based approach requires huge effort in adapting the verification setup for different configurations, which cannot meet the project timeline.
 
 \subsection{Detected Bugs}
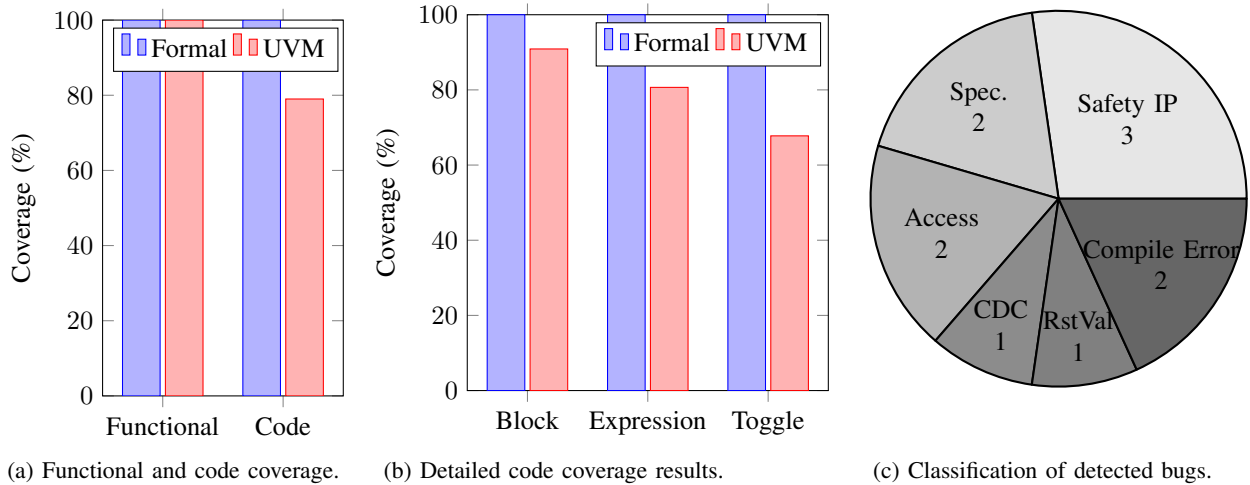
\begin{figure}

	\begin{subfigure}[b]{.3\textwidth}
		\centering
	\begin{tikzpicture}
		\begin{axis}[
			ymin=0,
			ymax=100,
			ybar,
			xtick=data,
			bar width=.5cm,
			ylabel={Coverage (\%)},
			xmin=0,
			xmax=4,
			x=0.8cm,
			y=0.05cm,
			xticklabels={Functional, Code},
			legend columns=2,
			ylabel near ticks,
			]
			\addplot table[x=interval,y=Formal]{\coveragegeneral};
			\addplot table[x=interval,y=UVM]{\coveragegeneral};
						\legend{Formal,UVM};
		\end{axis}
	\end{tikzpicture}
	\caption{Functional and code coverage.}
	\end{subfigure}
	\hspace{-4pt}
	\begin{subfigure}[b]{.3\textwidth}
		\centering
	\begin{tikzpicture}
		\begin{axis}[
			ymin=0,
			ymax=100,
			ybar,
			xtick=data,
			xticklabels={Block, Expression, Toggle},
			bar width=.5cm,
			ylabel={Coverage (\%)},
			x=0.8cm,
			y=0.05cm,
			xmin=0,
			xmax=6,	
			legend columns=2,		
			ylabel near ticks,
			]
			\addplot table[x=interval,y=Formal]{\coveragecode};
			\addplot table[x=interval,y=UVM]{\coveragecode};
			\legend{Formal,UVM};
		\end{axis}
	\end{tikzpicture}
	\caption{Detailed code coverage results.}
	\end{subfigure}
	\hfill
	\begin{subfigure}[b]{.3\textwidth}
	\begin{tikzpicture}
		\pie[color={black!10, black!20, black!30, black!45, black!50, black!60}, text = inside, sum=auto, after number=, radius=2.5]{3/Safety IP, 2/Spec., 2/Access, 1/CDC, 1/RstVal, 2/Compile Error}
	\end{tikzpicture}
	\vspace{8pt}
	\caption{Classification of detected bugs.}
	\end{subfigure}
	\caption{Verification results of the proposed framework.}\label{fig:results}
	\vspace{-10pt}
\end{figure}

 Finally, we would like to discuss the detected bugs, which were found in the generator verification. As mentioned before, given the specifications, all properties can be generated for the formal tools automatically and most of the human effort is spent in the collaboration with different teams, e.g. Project and Design Teams. The real time spent in the formal proving takes only a small portion. Therefore, it is possible to verify a large number of configurations in a small time when no alignments are required. Based on our results, eleven new design issues were found using this novel approach for the generator verified by the old approach, which demonstrate the completeness of formal approach. These detected bugs are classified in Fig.~\ref{fig:results}(c) and detailed explanations are presented as follows: 
 \begin{itemize}
 	\item Safety IP:  Three bugs are related with the Safety IP (SFF), where the connectivity of SFF instances violates the integration requirements.
 	
 	\item Specification (Spec.): We also found two issues in the specifications, which were documented wrongly.
 	
 	\item Access Policy (Access): Two bugs are found for the access policy. One is the access protection for one specific register is not implemented correctly and the other one is more serious where the bus error cannot be generated if the register kernels operate at multiple different clocks. 
 	
 	\item Clock Domain Crossing~(CDC): This issue was found using an industry standard clock domain crossing checking tool. When an asynchronous clock is used, there is a metastability issue for the synchronization block. 
 	
 	\item Reset Value (RstVal): In one special configuration, the reset values of read only registers do not match the specifications.
 	
 	\item Compile Error: In some corner case configurations, the generated RTL cannot be compiled due to syntax errors.
 \end{itemize}

\pgfplotstableread[row sep=\\,col sep=&]{
	interval    & Formal & UVM \\
	1  & 100    & 100 \\	
	3        & 100    & 79  \\ 
}\coveragegeneral

\pgfplotstableread[row sep=\\,col sep=&]{
	interval & Formal & UVM \\
	1       & 100  & 90.89  \\ 
	3  & 100  & 80.67  \\ 
	5      & 100  & 67.77  \\ 
}\coveragecode

\section{conclusion}\label{sec:conclusion}
The configurable feature of our in-house register generator brings high verification complexity for the verification engineer. In addition, two well-known issues make the register verification even harder. However, we have proposed an automated formal verification framework based on MDA, which can handle these issues and reduce the verification effort significantly. Moreover, this verification framework has better coverage compared with our UVM-based approach. In addition, more than ten new bugs were found in the generator verified by the old approach. Finally, our proposed formal framework has huge potential to support more customized features. 
\section*{Acknowledgment}
We also want to thank our colleagues, Vidya Sagar Kantamneni, Yifang Wang, Xin An and Djones Lettnin, for their valuable input and suggestions.

\bibliographystyle{IEEEtran}
\bibliography{bibfile}
\end{document}